# Analysis of the Robustness of Conventional and Topologically Protected Edge States in Phononic Crystal Plates

Yabin Jin[1,*], Daniel Torrent[2,*], Bahram Djafari-Rouhani[3]

[1]Institut de Mécanique et d'Ingénierie, UMR CNRS 5295, Université de Bordeaux, 33405 Talence, France
[2]GROC-UJI, Institut de Noves Tecnologies de la Imatge, Universitat Jaume I, 12080, Castello, Spain
[3]Institut d'Electronique, de Microélectonique et de Nanotechnologie, UMR CNRS 8520, Département de Physique, Université de Lille, 59650 Villeneuve d'Ascq, France
[*]Corresponding authors: yabin.jin@u-bordeaux.fr;dtorrent@uji.es

In this work we theoretically study the interface acoustic states of resonators on a thin plate with topologically protected and conventional designs. Topologically protected interface state is first analyzed by employing the conception of breaking inversion symmetry within the unit cell of a honeycomb lattice for cylindrical and spherical resonators; we further demonstrate the robustness of the wave propagation along a zig-zag path containing sharp corners, defect and disorder. The wave propagation ceases to be preserved if we increase the degree of disorder along the zig-zag path. In parallel, the conventional interface state is also designed and compared to the same situations. We found that the conventional interface state suffers back scattering in the zig-zag path while it can show a more confined wave transport in some cases. The presence of a defect along the propagation path scatters the conventional interface wave and in particular can prohibit a full propagation in presence of a localized state at the defect. If the zig-zag path is made disordered, the propagation of the conventional interface mode can be conserved at given frequencies for a low random degree and disappears for higher random degree as the interface bands become flat in dispersion and turn to localized states. Finally, we show that the immunity of the topologically protected design needs the interface to be surrounded by at least two hexagons of the phononic crystals on both sides, especially at the sharp corners in the zig-zag path, while the conventional design only needs one hexagon bulk media with the advantage of compact wave transport. This work puts a step forward for the interface states in micro-/nano-scale characterization and figures out the behaviors for both topologically protected and conventional interface states.

## 1. Introduction

Phononic crystal plates[1-3], which consist of periodic resonators or scatters attached to an elastic plate, have received increasing attention from the past decade. These structures can exhibit hybridization and Bragg band gaps simultaneously, so that they present aspects of both phononic crystals [4-8] and acoustic metamaterials[9-12]. In a particular case for flexural wave, analytical models describe a pillar as a point-like resonator connecting with the plate by a string with out of plane displacement component motion, which found topologically protected edge states recently[13, 14]. It should be noted that topological structures with realistic spheres have not been studied until now and only few works reported with realistic pillars[15, 16]. How the topologically protected edge states behave with different resonators has not been studied yet.

Additionally, these realistic structures have (several) dispersion curves besides those at the Dirac point and here we show that the topological effect for flexural waves can be preserved despite these other dispersion curves. When the resonators are organized in a honeycomb arrangement, the dispersion curves of flexural waves present the so called Dirac points, and bounded edge states can be found for finite slabs [17]. Then by breaking inversion symmetry in the unit cell of the honeycomb arrangement, topologically nontrivial band gaps are found which support topologically protected edge states [13, 15, 16, 18].

Topologically protected edge states were originally found in electronic systems [19], and their mechanical analogues have been widely studied, like the analogue to the quantum Hall effect[20], the quantum spin Hall effect[21-23] and the quantum valley Hall effect[24]. All these guided states are protected against local defects and sharp bends, which means that waves excited along one direction are free of being backscattered by disorder or defects. Edge states however are not unique of topological insulators, since whenever there is an interface, there is the possibility of finding guided states at this interface, and a comprehensive comparison between topologically protected states and "conventional" edge states has not been reported so far for mechanical waves.

In this work, we will study the topologically protected edge states in a thin elastic plate with cylindrical or spherical resonators attached to it. We will design a plate supporting both topologically protected and conventional edges states, and we will analyze their robustness for propagation along a zig-zag path and under the presence of defects and disorder in the zig-zag edge. Also, we will analyze the size of the bulk material surrounding the interface to continue supporting this type of state. The paper is organized as follows: after the introduction, Section 2 will analyze the dispersion curves of plates with attached spherical and cylindrical resonators in honeycomb lattice. In Section 3 we will design topologically protected edge states with both types of resonators by employing two different resonators in the unit cell, in order to break the inversion symmetry, and we will analyze its robustness under local perturbations. We will also analyze the minimum size of bulk media surrounding the interface to preserve the existence of the edge state. In Section 4 we will design a similar plate supporting conventional edge states and we will compare its properties with those of the topologically protected edge state. Finally the work is summarized in Section 5.

## 2. Spherical and cylindrical resonators over a thin elastic plate

We consider a honeycomb arrangement of resonators (Fig. 1*a*) attached to a thin elastic plate. Two types of resonators are considered in this work: a cylindrical pillar deposited over the plate (Fig. 1*b*) and a spherical scatter attached to the plate by means of a very small contact area (Fig. 1*c*). These two types of resonators are most feasible from the practical point of view, especially at the micro- or nano-scale, than other approaches based on inclusions in the plate.

In the numerical simulations performed in this work by means of Finite Element Method (FEM), the sphere and cylinder resonators are defined as fused silica[25], with elastic modulus $E_s=E_c=62$GPa, Poisson's ratio $\nu_s=\nu_c=0.24$, and mass density $\rho_s=\rho_c=2200$kgm$^{-3}$ where the subscripts *s* and *c* stands for sphere and cylinder, respectively. The thin plate is made of aluminum, whose elastic parameters are $E_p=73$GPa, $\nu_p=0.17$ and $\rho_p=2730$kgm$^{-3}$ where the subscript *p* stands for plate. It is also useful to define the normalized frequency Ω*a* as

$$\Omega^2 = \omega^2 \rho_p a^2 e / D \tag{1}$$

where $D = E_p e^3 / 12(1-\nu_p^2)$ is the plate's bending stiffness, *a* is lattice constant, and *e* is the thickness of the plate.

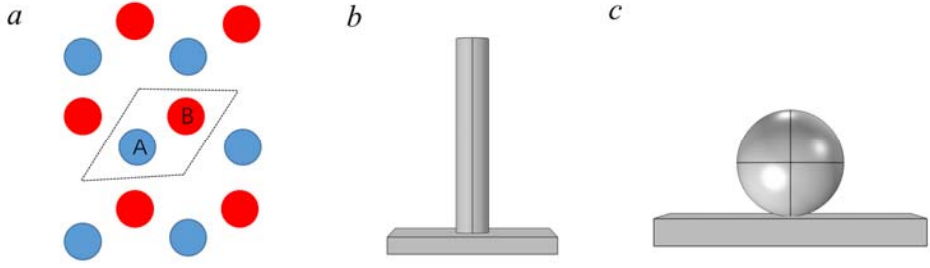

Figure 1: Phononic crystal plate with honeycomb lattice (*a*); cylinder (*b*) and sphere (*c*) resonators are deposited on a thin plate.

Figure 2*a* shows the band structure computed by the Finite Element Method (FEM) of a honeycomb lattice of cylindrical resonators attached to the elastic plate ("pillared" phononic crystal plate). The radius of the cylinders is $R_c$=0.1*a* and their height $h_c$=1.2*a*. The right panel shows the same lattice but with spherical resonators ("sphered" phononic crystal plate), where the spheres have the same volume as the cylinders, which results in a radius $R_s$=0.208*a*. In both cases the plate's thickness is chosen as *e*=0.1*a*. It is found that for Ω*a*<5 the anti-symmetric Lamb mode, identified by its parabolic trend in the low frequency limit, interacts with the local resonant modes of the scatters (bending resonance of the cylinder, bending and rotating resonances of the sphere), then hybridization band gaps appear as well as nearly flat branches.

The degeneracy created by the existence of two resonators per unit cell results in the emergence of the so-called Dirac cone at the *K* point of the dispersion curve, as can be seen in Fig.2*a* and *b* for cylindrical and spherical resonators, respectively. The existence of Dirac cones in these lattices have been widely studied in photonics[26, 27] and acoustics[28, 29], with remarkable properties for the propagation of waves in the vicinity of the vertex of the cones. Special mention deserves the recent rise of the domain of topological insulators[19], where topologically protected edge states with one-way propagating properties can be properly tailored on the basis of these states[13, 15]. However, for elastic plates these states have been found in idealized systems of point-like resonators[13, 14], whose physical implementation has been scarcely discussed in the literature. Although other complex systems have been implemented at the macro-scale[15], their realization in the micro or nano-scale has not been properly studied yet. Therefore, the existence of Dirac cones in the structures proposed before, more suitable for the micro or nano-scale, opens the door to a new type of devices based on the extraordinary properties of topological insulators.

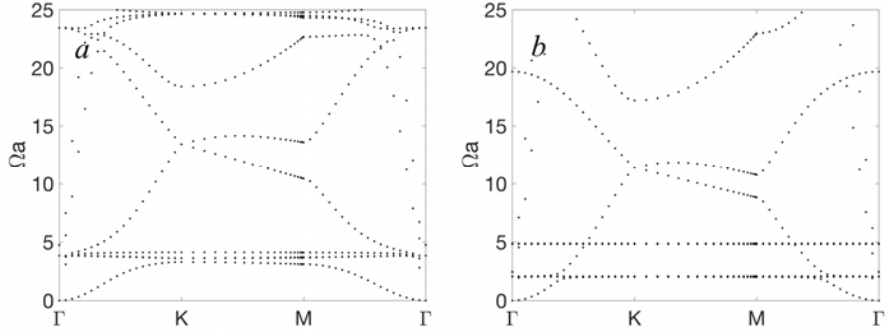

Figure 2: Dispersion of pillared (*a*) and sphered (*b*) phononic crystal plate with honeycomb lattice arrangement with identical scatters A and B. Geometric parameters are lattice parameter *a*, thickness of the plate *e*=0.1*a*, radius of the pillar $r_p$=0.1*a*, height of the pillar *h*=1.2*a*, radius of the sphere $r_s$=0.208*a*. The pillar and the sphere have a same volume.

To be noted that the diameter and the height of the cylindrical resonator are chosen to well isolate the Dirac point from other resonant flat bands at K point. From Fig.2*a*, the Dirac point locates more or less at the middle of the first (about Ω*a*=5) and the second (about Ω*a*=25) bending modes of the cylindrical resonators, so that these modes does not interact with the topological modes of interest.

## 3. Topologically protected interface states

In the honeycomb lattice the mirror symmetry is broken when the two resonators A and B of the unit cell are different. In the dispersion curves depicted in Fig.3 the mass of resonator A is 3 times that of resonator B, $m_A$=3$m_B$. It is clear that a band gap for the flexural mode is opened at the Dirac cone for both cylindrical and spherical resonators shown in Fig.2. It is well known that the Chern number for the upper and lower bands with respect to the gap have opposite signs, what makes these systems suitable to support topologically protected states in this non-trivial band gap[13, 15], resulting from the elastic analogue of quantum valley Hall effect. Although this non-trivial band gap is only for flexural waves, it is possible to chose the parameters of the crystals such that the flexural wave remains almost independent from symmetric and shear-horizontal waves[30] and the topologically protected states for flexural wave works[15, 16].

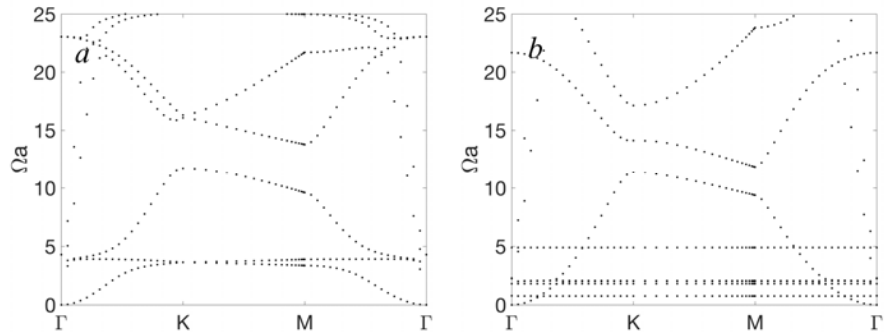

Figure 3: Dispersion of pillared (*a*) and sphered (*b*) phononic crystal plate with honeycomb lattice arrangement with asymmetric resonators A and B. Geometric parameters are lattice parameter *a*, thickness of the plate $e=0.1a$, radius of the pillar $r_p=0.1a$, height of the pillar $h_A=1.2a$ and $h_B=0.4a$, radius of the sphere $r_{sA}=0.208a$ and $r_{sB}=0.144a$.

Figure 4 shows the geometry considered for the analysis of these states, where a stripe consisting of two lattices, B-A and A-B, is considered. The two "A" resonators at the interface are marked by the black dotted rectangle at the bottom of the figure. As before, cylindrical (Fig.4*a*) and spherical (Fig.4*b*) scatters are considered. The stripe is assumed to be periodic in the *x* direction. From the dispersion diagram, we find an edge mode traversing the opened band gap, which is highlighted in bright color and limited by the dotted rectangle domains. The color bar in the dispersion indicates a displacement ratio $\alpha$ between the two interface-A-A resonators and the whole stripe, as[31]

$$\alpha = \frac{\iiint_{A-A} \sqrt{u_x^2+u_y^2+u_z^2}\,dv}{\iiint_{Stripe} \sqrt{u_x^2+u_y^2+u_z^2}\,dv} \qquad (2)$$

The non-trivial gap for the cylindrical resonator occupies the normalized frequency domain (11.5, 12.8), three times wider than that of the spherical one (11.0, 11.5) and results in a higher $\alpha$ value in wider band-gap as shown in the left panel of cylinder system.

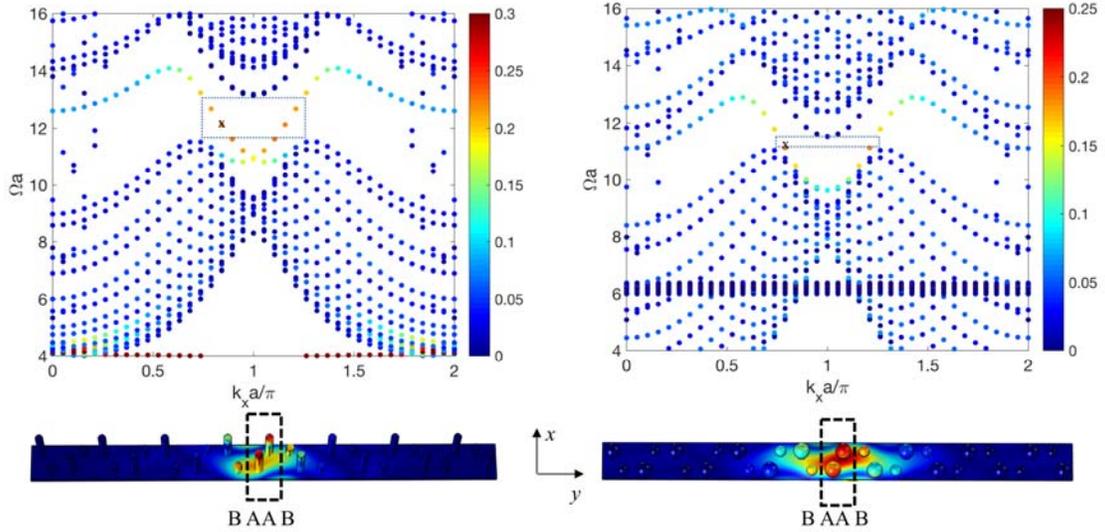

Figure 4: Dispersion of a stripe consisting B-A and A-B honeycomb lattices with the interface as B-A-A-B. The color in the dispersion represents the ratio of total displacement localized in the A-A resonators (black dotted rectangle) over the whole stripe. The displacement distributions at marked 'x' point in the dispersion are shown at the corresponding bottom.

To demonstrate the topological robustness of this edge state, we construct a zig-zag channel for the B-A-A-B type interface on a circular thin plate. A point-like flexural source (finite line source with the length as *a*) is excited at the bottom exit of the zig-zag path. A topologically protected waveguide is able to suppress backscattering and robust transport of phonons along the zig-zag interface, which are clearly demonstrated for cylindered (left panel) and sphered (right panel) resonators in Fig.5. As explained in Fig.4, the cylindered resonator system has a wider non-trivial band gap so that the phonon transport is more concentrated at the interface, with less leakage in the bulk. In Fig.6, we remove a resonator at the middle interface as pointed by the red arrow, the same interface transports are found for both cylindered and sphered resonators, behaving as robust wave propagation after the imperfection.

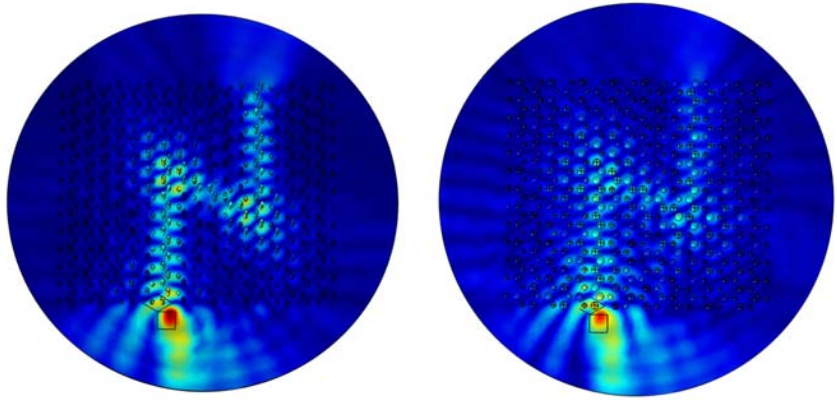

Figure 5: Topologically protected edge state is excited by anti-symmetric Lamb wave at $\Omega a= 11.98$ for pillar-resonator (left panel) and at $\Omega a=11.31$ for sphere-resonator (right panel) with backscattering-immune property.

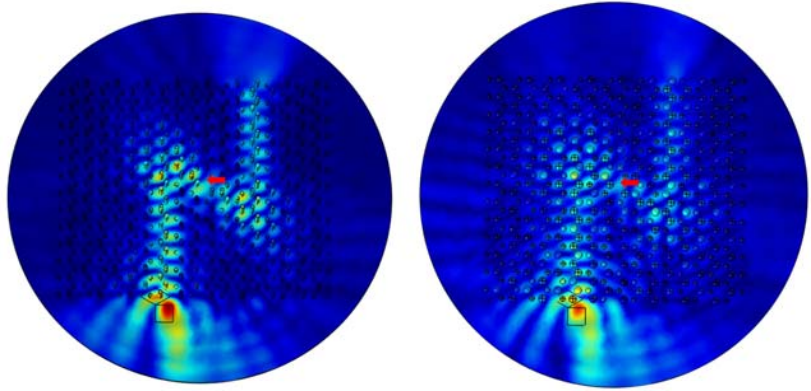

Figure 6: Topologically protected edge state is excited at $\Omega a= 11.98$ for pillar-scatter (left panel) and at $\Omega a=11.31$ for sphere-scatter (right panel) when one resonator (pointed with red arrow) is removed

Due to the smaller band gap in the case of spherical resonators, and hence a less confinement of the interface mode, we limit the following discussions to the case of cylindrical pillars.

In previous studies[18, 21, 22], most works have reported that the topologically protected edge state can support local disorder in the edge path. However, up to now, the influence of the disorder has been mainly limited to a small part of the path and disorder along the whole edge path is rarely reported. In specific chiral hyperbolic metamaterial with a photonic Weyl system, the disorder-induced state transition was found recently in such photonic topological system[32].

Here, we allow all the pillars $i$ at the interface to be randomly displaced $l_i=(x_i, y_i)$ with respect to their original positions, as shown by the red dots in Fig.7$a$ and $b$. The disorder degree $\gamma$ is defined as

$$\gamma = \max( \ |l_i| \ ) \tag{3}$$

which means the maximum deviation (big black circle in Fig.7$a$ or $b$) of pillars from their original positions in the zig-zag interface. In Fig.7$a$ and $b$, $\gamma$ is defined as $0.1a$ and $0.2a$, respectively, as shown by the black solid circles containing the random positions of the pillars along the interface path. For instance, for $\gamma=0.1a$ the moved displacement of pillar $i$ in the interface is $0.06a$ only in $x$ axis, corresponding to the red dot $l_i=(0.06a, 0)$ in Fig.7$a$; for $\gamma=0.2a$ the moved displacement of pillar $j$ in the interface is $-0.17a$ in $x$ axis and $0.08a$ in $y$ axis, corresponding to the red dot $l_j=(-0.17a, 0.08a)$ in Fig.7$b$. Once all the pillars along the zig-zag interface are moved with random displacements corresponding to the red dots, a random interface is constructed.

Figures 7$c$ and $d$ show the numerical simulations of the wave propagation along the zig-zag path at $\Omega a= 11.98$ for $\gamma=0.1a$ and $\gamma=0.2a$, respectively, from which one can observe that the robustness of the topologically protected edge state is still conserved at $\gamma=0.1a$. However, at $\gamma=0.2a$, elastic waves are localized at several positions in the zig-zag path, showing the existence of back-scattered waves which form standing waves along the interface, and the breaking of the robustness.

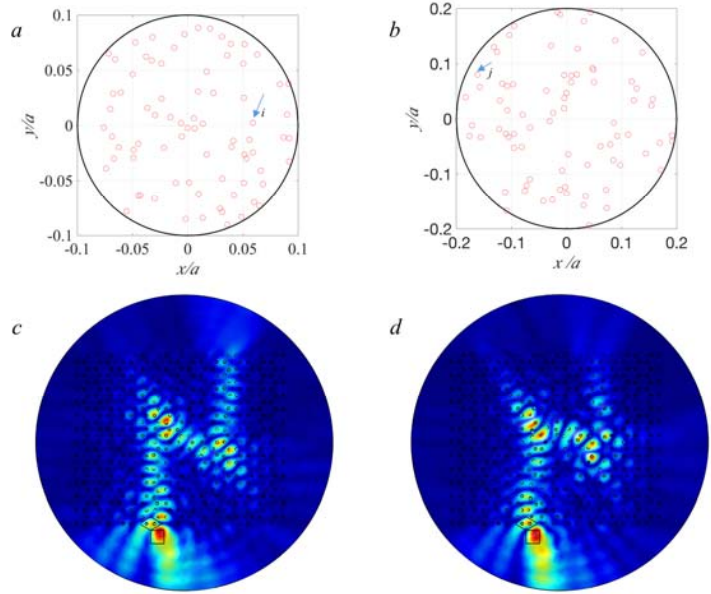

Figure 7: The random degree $\gamma$ for the pillars along the zig-zag interface $\gamma=0.1a$ (*a*) and $\gamma=0.2a$ (*b*). The red circles stand for the random moved displacement for any pillar in the zig-zag interface path. Simulations at $\Omega a= 11.98$ for $\gamma=0.1a$ (*c*) and $\gamma=0.2a$ (*d*). Red dots $i$ and $j$ are pointed out for examples.

The topologically protected interface mode relies on the emergence of the non-trivial band gap from the bulk phononic crystal, which obviously requires the size of the bulk media to be sufficiently large. In the previous examples, the length of the bulk material was at least four unit cells on each side of the zig-zag interface. Figure 8 shows the same simulations when the bulk media is reduced to only one hexagon on each side of the vertical straight parts of the interfaces (panel *a*) and then to one hexagon on any side of the zig-zag interface (panel *b*). Panel *a* shows that, with the minimized bulk phononic crystal, the effect of leaky wave is strengthened and the propagating state in the upper straight interface part decreases. Nevertheless, the zig-zag interface state is still visible. However, when the bulk media is further minimized as shown in panel *b*, flexural waves are transported along the lower straight interface part and exit to the background plate directly, so that the flexural vibrations do not enter the middle and the upper parts of the zig-zag interface. Therefore, one hexagon of bulk media can be the minimum size for straight interface state and at least two hexagons of bulk media is needed for zig-zag interface especially around the sharp corners.

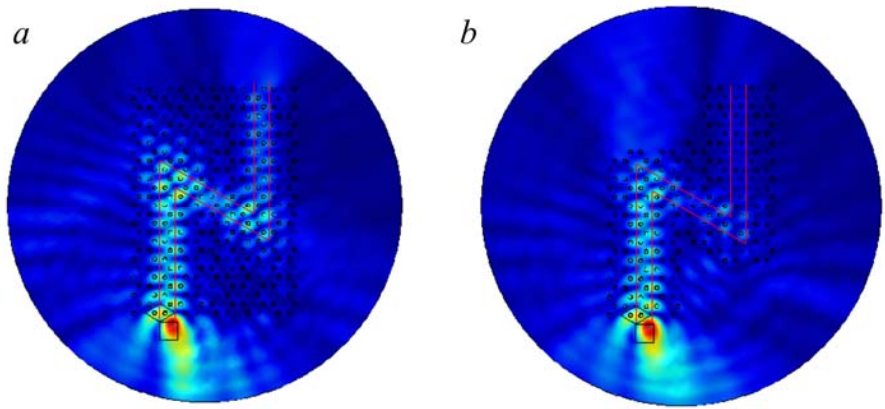

Figure 8: Minimized bulk size with one hexagon unit on one side of the straight interface part (*a*) or on any side of the zig-zag interface (*b*). Flexural wave source is excited at $\Omega a = 11.98$.

## 4. Conventional interface states

An interface state can also be found for a zig-zag type honeycomb lattice edge with different path shapes[17]. In this section, we present a new way to design edge state by band engineering with the displacement concentration formula (2). Let us consider a honeycomb unit cell with two cylindrical pillars of the same radius $R_c=0.2a$ but different heights $h_A=0.8a$ and $h_B=0.4a$. The thickness of the plate is kept to $e=0.1a$. Notice that comparing to the parameters in Section 3, the radius increases and the height decreases, which contribute to increase the frequency of pillar's bending and compressional resonances[33], especially the blue shifted compressional resonance moves the potential Dirac point to a higher frequency range $\Omega a>25$ beyond the first two resonant modes. In this section, the conventional edge state design does not depend on the Dirac point, but directly on the dispersion tailoring of a finite stripe.

We first construct a finite stripe with one unit as shown in the lower part of Fig.9a. The stripe has a finite size along the *y* axis and Bloch-periodic boundary conditions are applied at the two edges along *x* direction. The dispersion is calculated and showed with the color standing for the displacement concentration rate of the four pillars at the two edges in *y* axis. Two bright and flat modes appear near the frequencies $\Omega a=12$ and $\Omega a=23$. The eigenmodes at $k_x a/\pi=1$ of these two modes, labeled k1 and k2, are shown at the bottom in the left panel, revealing localized states of this finite stripe in one edge (pillar B at the left edge) and not at the other (pillar A at the right edge). Then we take the pillar B to design an interface as type A-B-B-A in the middle of a stripe, shown as the black dotted rectangle at the bottom of Fig.9*b*. In the corresponding dispersion curve in Fig.9*b*, the color stands for the displacement concentration rate of the four pillars at the interface (in the black dotted rectangle). Comparing Fig.9*a* and *b*, two new bands appears as localized interface modes in Fig.9*b*, where their eigenmodes of k11 and k22 are displayed. From the color bar maximum limit and the interface eigenmodes, the conventional interface states are more confined than the topologically protected design in Fig.4.

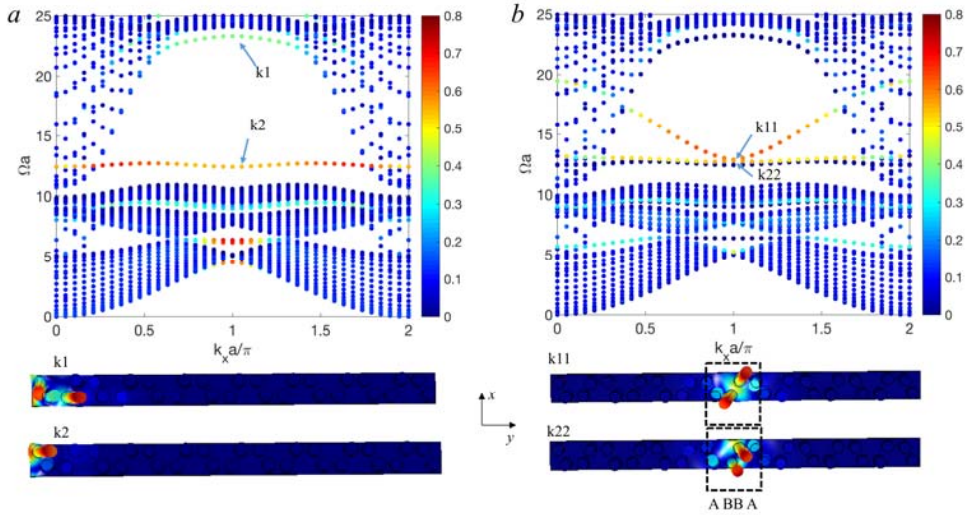

Figure 9: (*a*) Dispersion of a stripe consisting unit cell with one honeycomb lattice as $h_A$=0.8*a* and $h_B$=0.4*a*. The edge states at k1 and k2 points shown at the bottom. The color stands for the displacement concentration rate of the four pillars at the two edges in *y* axis (*b*) Dispersion of a stripe consisting unit cell with two lattices as an interface type ABBA in the middle. The color stands for the displacement concentration rate of the four pillars at middle interface as marked with black dotted rectangle. The interface states at k11 and k22 points shown at the bottom. Thickness of the plate *e*=0.1*a*, radius of the pillar $r_p$=0.2*a*.

A zig-zag A-B-B-A type interface with sharp corners is designed as in the previous section in order to compare the robustness of the interface states. The same point-like flexural wave source is excited at the exit of the bottom of the interface at $\Omega a$=13.30 and $\Omega a$=17.74 in Fig.10*a* and *b*, respectively, showing the conventional interface mode works in a broadband frequency region. Nevertheless, we observe that the amplitude of the flexural wave propagation decreases when passing through the zig-zag interface due to the back-scattering effect at the sharp corners, while it almost remains uniform for the topologically protected edge state in Fig.5*a*. It is important to remark that the conventional edge mode is more "compact", in the sense that the evanescence of the wave into the surrounding bulk is faster.

To better understand the behavior of the interface states and the propagation along a zig-zag path with defect and disorder, a $4a \times 4.3a$ model consisting B-B pillars interface in the dotted rectangle separated by one hexagon in *y* direction is built and Bloch-periodic boundary conditions are applied to the edges in both *x* and *y* directions, as shown in Fig.10*c*. The dispersion curves along the ΓX direction are displayed in Fig.10*d,* with the color standing for the displacement concentration rate for 8 pillars in the dotted rectangle over the whole model. It's clear that an interface mode (folded green bands) appears in the frequency range Ω*a* =[12.66, 16], which is in coincidence with the band associated to k11 in Fig.9*b*. Similarly, the band k22 in Fig. 9(b) gives rise to a set of flat localized interface modes around Ω*a* =12.6. The eigenmodes of points m1 and m2 at Γ point are displayed in Fig.10*e* and *f,* respectively, showing a compact interface propagation. Similar compact interface state is also found for k22 mode.

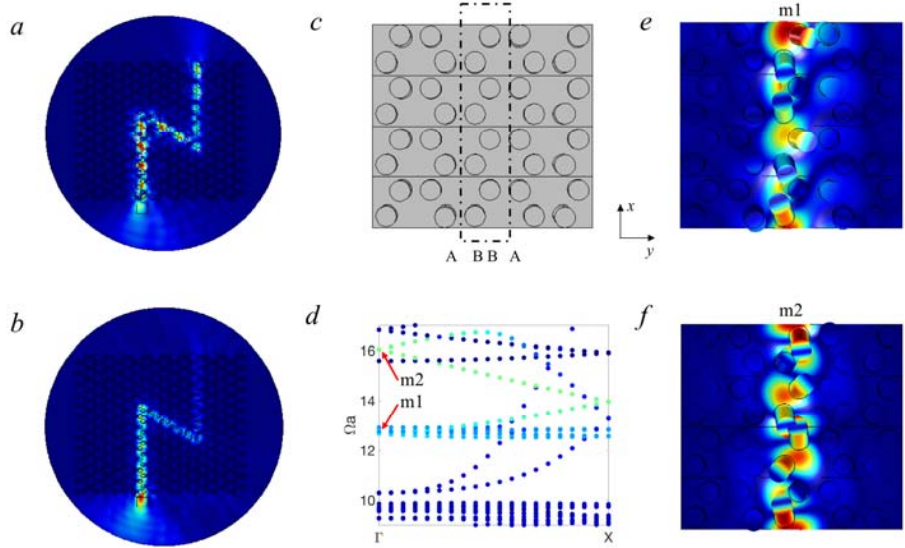

Figure 10: Conventional interface states at Ω*a*= 13.30 (*a*) and Ω*a*= 17.74 (*b*) with the interface ABBA type in Fig.9*b*; (*c*) Geometry of a super-cell model with two lattices as A-B-B-A type interface (dotted rectangle) in the middle to better understand the conventional interface mode. The model has a size of 4*a* in *x* and 4.33*a* in *y*, with Bloch-periodic conditions applied to the edges in both directions The dispersion along ΓX is calculated and shown in (*d*) with two points at Γ point as m1 and m2; The eigenmodes of m1 and m2 points are shown in (*e*) and (*f*), respectively.

In order to check the robustness of the propagation for the conventional edge state, a pillar at the same position in the zig-zag path as in Fig.6 is removed to see how the mode interacts with the defect. In Fig.11 upper panel, simulations performed at Ω*a*= 13.30 (*a*), Ω*a*= 13.75 (*b*) and Ω*a*= 16.41 (*c*) are presented as back-scattered, defect localization and propagation phenomena, respectively. We again calculate the dispersion curves of the 4*a* × 4.3*a* model (Fig. 10*c*) but containing now the defect, which consist of the removal of one pillar. The result along the ΓX direction is shown in Fig. 11*d* (the color stands for the displacement concentration rate on the two pillars in the neighborhood of the absent pillar position in the dotted rectangle). We find that the insertion of the defect will upward shift the folded interface mode, starting now at Ω*a*= 13.75 at the Γ point. The eigenmode at this Ω*a*= 13.75 point is shown in (*e*) with a defect localized mode[34]. This band gradually changes from defect mode (Ω*a*= 13.75) to a normal interface mode (Ω*a*= 16). The eigenmode of frequency Ω*a*= 16 at Γ point is shown in panel (*f*). The up-shift defect-interface branch results in a very narrow band gap. The operating frequency $\Omega a=$ 13.30 locates inside the narrow band gap while also close to several flat localized modes, so that it is partial propagating state in the lower straight waveguide of the zig-zag path with back scattering effect. The narrow band gap, defect mode at $\Omega a=$ 13.75 and interface propagating mode at $\Omega a=$ 16 explain the different behaviors in (*a*), (*b*) and (*c*), respectively**.** Meanwhile for topologically protected interface mode in Fig.6, the propagation is protected against the defect as in Fig.5.

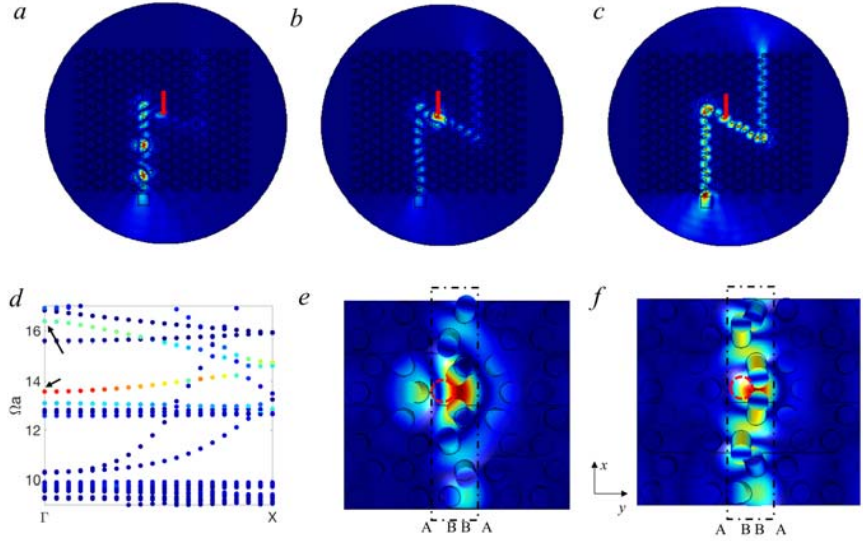

Figure 11: Conventional interface states at $\Omega a=$ 13.30 (*a*), $\Omega a=$ 13.75 (*b*) and $\Omega a=$ 16.41 (*c*) when moving a pillar in the central of zig-zap path (pointed with red arrow) as in Fig.9*a*; (*d*)dispersion along ΓX when moving one pillar in the interface in Fig.8*c*; The eigenmodes of $\Omega a=$ 13.75 (*e*) and $\Omega a=$ 16.41 (*f*) at Γ point marked with black arrow in (d) are shown, with the absent pillar indicated as the red dotted circle.

In Fig.12, random degrees $\gamma=0.1a$ and $\gamma=0.2a$ are introduced for 8 pillars in the dotted rectangle of the super-cell unit cell used in Fig.10c, and the dispersion curves are computed along the ΓX direction. The upper panels (*a*, *b*, *c*) refer to the low disordering $\gamma=0.1a$ whereas the lower panels correspond to the high disordering $\gamma=0.2a$. Each row in Fig. 12 gives respectively the propagation at the zig-zag interface path, the dispersion curves of the super-cell of Fig. 10*c*, and the localization of the interface modes at two chosen frequencies. For random degree $\gamma=0.1a$, the interface state can still be found as the example in (*a*) with $\Omega a=$ 17.7. From dispersion in (*b*) (8 pillars are randomly moved for $\gamma=0.1a$), the interface mode is now divided into several isolated bands and each band becomes more flat. The eigenmode of $\Omega a=$ 16.41 at Γ point is shown in (*c*), supporting a conserved interface state. To be noted that the 8 random pillars in the dotted rectangle for eigenmode calculation in Fig. 12 (*c*) is much less than the real propagation in Fig. 12 (*a*) and different random distributions can make the interface mode branch interacting with the local defects behaving differently, e.g. the narrow band gap appears at different frequencies with different widths. Therefore it is possible that the real operating frequency $\Omega a=$ 17.7 in full wave propagation is not the same as the eigenmode $\Omega a=$ 16.41 in the particular dispersion calculation of Fig. 12 (*c*).

When the random degree increases to $\gamma=0.2a$, the dispersion in (*e*) (8 pillars are randomly moved for $\gamma=0.2a$) shows more divided isolated bands becoming more flat, turning to be random localized modes, as the example of $\Omega a= 16.23$ at $\Gamma$ point shown in (*f*) the displacements are not uniformly distributed as in (*c*). Therefore, the random degree $\gamma=0.2a$ can't support interface mode any more. Full simulation at $\Omega a= 16.23$ is displayed in (*d*), showing that the flexural wave can't propagate with the first sharp corner with back-scattering effect and waves are localized near the sharp corner. In full simulation, it can be affected by both flat branch and the narrow band gap at this operating frequency $\Omega a= 16.23$, resulting in back-scattering effect. Prohibited propagation can also be found for specific frequencies located at the middle of the band gap.

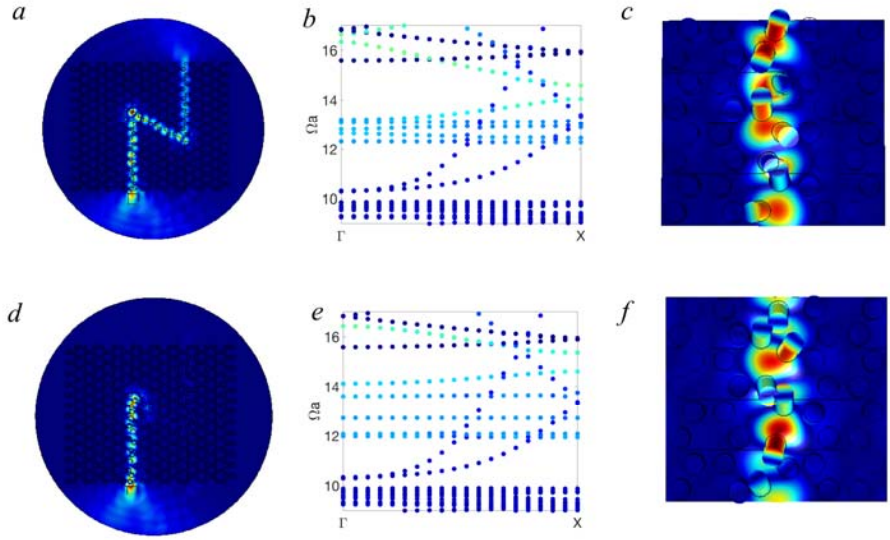

Figure 12: Upper panel for random degree $\gamma=0.1a$: (*a*) Zig-zag interface state excited at $\Omega a= 17.7$ with conventional design in Fig.11(*b*); (*b*) $\Gamma X$ dispersion when $\gamma=0.1a$ is applied to all pillars in the dotted rectangle; (*c*) eigenmode of the interface state $\Omega a= 16.41$ at $\Gamma$ point; Lower panel for random degree $\gamma=0.2a$: (*d*) Zig-zag interface state excited at $\Omega a= 16.23$ with conventional design; (*e*) $\Gamma X$ dispersion when $\gamma=0.2a$ is applied to all pillars in the dotted rectangle; (*f*) eigenmode of the interface state $\Omega a= 16.23$ at $\Gamma$ point;

Finally, we study the minimum size of the bulk phononic crystals surrounding the straight or zig-zag interfaces to allow the interface state propagation. Unlike the minimized bulk phononic crystals size for topologically protected interface state in Fig. 8, we show in Fig.13 that the conventional interface states are conserved for such a minimized surrounding of the interface, namely one hexagon bulk media surrounding the sharp corner. This result brings also an a posteriori support to the validity of the model in Fig.10*c*, which was used to studying the propagation of interface state in the zig-zag path in presence of defects and disorder. Indeed, in this model, the interface has exactly the same surroundings as in Fig. 13*b*.

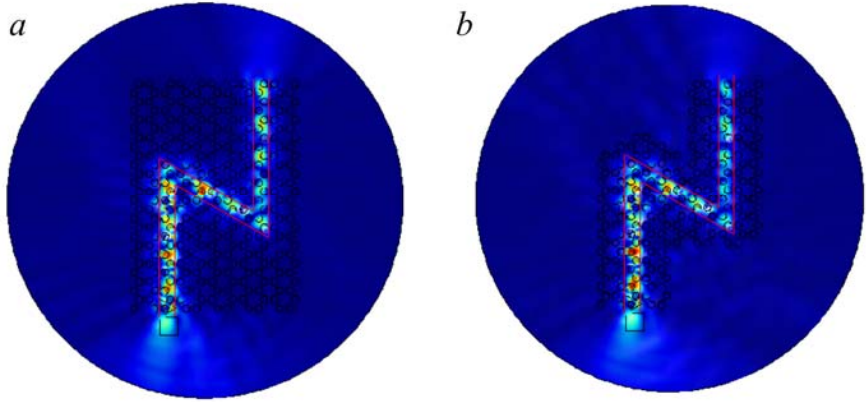

Figure 13: minimized bulk size with one hexagon unit on one side of the straight interface part (*a*) or on any side of the zig-zag interface (*b*). Flexural wave source is excited at $\Omega a$= 13.30.

## 5. Summary

This work analyzed the dispersion of both cylindrical and spherical resonators on a thin plate in honeycomb lattice, and we employed the known conception of breaking the inversion symmetry in the unit cell to excite topologically protected edge states. These are found for both types of resonators, being more robust in the cylindrical case. We compared their properties with those of a conventional interface state without topological protection, which exhibits more confined propagation along a zig-zag path. Due to the back-scattering effect at the sharp corner of the path, the propagating wave's amplitude of the conventional edge state decreases when passing through the corners, unlike the topological states which are protected against back-scattering. We also compared the behaviors of conventional and topologically protected interface states with the presence of defects and disorder. The topologically protected interface state is immune to the defect, while the conventional interface state changes to back scattered, defect localized and propagating states resulting from the interaction between the defect and the interface mode. Random perturbations were added also to the zig-zag edge, and it was shown that topologically protected state are in general more robust than conventional states. We also analyzed the required size of the bulk media to conserve the edge states, showing that conventional states require less bulk material than topologically protected states. These comparisons help to understand, distinguish and define conventional or topologically protected states, whose right understanding is more than relevant to properly develop the potential applications of these devices.

**Acknowledgement**
Work supported by the U.S. Office of Naval Research under Grant No. N00014-17-1-2445. D.T. acknowledges financial support through the "Ramón y Cajal" fellowship.